\documentclass[12pt]{article}
\usepackage{amsmath}
\usepackage{epsfig}
\usepackage{amstext,amssymb,amsfonts}

\advance\voffset by -1.5cm
\advance\hoffset by -1.25cm
\def\Tr{\,{\rm Tr}\, }

\def\Im{\,{\rm Im}\, }

\def\be{\begin{equation}}
\def\ee{\end{equation}}
\def\ba{\begin{eqnarray}}
\def\ea{\end{eqnarray}}

\newcommand{\eg}{{\it e.g.~}}
\newcommand{\ie}{{\it i.e.~}}

\newcommand{\Zop}{\mathbb{Z}}

\newcommand{\nn}{{\nonumber}}

\begin{document}

\vspace*{-1.5cm}
\thispagestyle{empty}
\begin{flushright}
hep-th/0609034
\end{flushright}
\vspace*{2.5cm}

\begin{center}
{\Large 
{\bf Bulk induced boundary perturbations}}
\vspace{2.5cm}

{\large Stefan Fredenhagen}%
\footnote{{\tt E-mail: stefan@itp.phys.ethz.ch}},
{\large Matthias R.\ Gaberdiel}%
\footnote{{\tt E-mail: gaberdiel@itp.phys.ethz.ch}} 
{\large and} 
{\large Christoph A.\ Keller}%
\footnote{{\tt E-mail: kellerc@itp.phys.ethz.ch}} 
\vspace*{0.5cm}

Institut f{\"u}r Theoretische Physik, ETH Z{\"u}rich\\
CH-8093 Z{\"u}rich, Switzerland\\
\vspace*{3cm}

{\bf Abstract}

\end{center}

\noindent 
The influence of closed string moduli on the D-brane moduli space is 
studied from a worldsheet point of view. Whenever a D-brane cannot be
adjusted to an infinitesimal change of the closed string background,
the corresponding exactly marginal bulk operator ceases to be exactly
marginal in the presence of the brane. The bulk perturbation then
induces a renormalisation group flow on the boundary whose end-point
describes a conformal D-brane of the perturbed theory. We
derive the relevant RG equations in general and illustrate the
phenomenon with a number of examples, in particular the radius
deformation of a free boson on a circle. At the self-dual radius 
we can give closed formulae for the induced boundary flows which are
exact in the boundary coupling constants.   
\newpage
\renewcommand{\theequation}{\arabic{section}.\arabic{equation}}

%%%%%%%%%%%%%%%%%%%%%%%%%%%%%%%%%%%%%%%%

\section{Introduction}

The problem of how to stabilise the moduli of phenomenologically
interesting string backgrounds is currently one of the central
questions in string theory (for recent reviews
see~\cite{Douglas:2006es,Blumenhagen:2006ci}). Most backgrounds of
interest involve D-branes, and thus there are two kinds of moduli to
consider: the D-brane moduli that describe the different D-brane
configurations in a given closed string background, and the closed
string moduli that characterise this closed string
background. Obviously, these two moduli spaces are not independent of
one another: the moduli space of D-branes depends on the closed string
background, and thus on the closed string moduli. On the other hand,
the D-branes `back-react' on the background, and thereby modify the
original closed string background in which they were placed. In order
to make progress with stabilising all moduli in string theory, it is
therefore of some significance to understand the interplay between
these two moduli spaces better.  \smallskip

In this paper we make a small step towards this goal. It is well known
that the closed string moduli space is described, in conformal field
theory, by the exactly marginal bulk perturbations. A necessary
condition for a bulk field to be exactly marginal is that it has 
conformal weight $(1,1)$, and that its three-point self-coupling
vanishes \cite{Kadanoff:1978pv,Cardy}. This condition was derived for
conformal field theories without boundary, but in the presence of a
D-brane, the situation changes. Indeed, a marginal bulk operator that
is exactly marginal in the bulk theory may cease to be exactly
marginal in the presence of a boundary. 

The simplest example where this phenomenon occurs, is the theory of a
single free boson compactified on a circle. For this theory the full
moduli space of conformal D-branes is known
\cite{Gaberdiel:2001zq,Janik} 
(see also \cite{Friedan,Recknagel:1998ih}). It depends in
a very discontinuous manner on the radius of the circle, which is one
of the bulk moduli. We always have the usual Dirichlet and Neumann
branes, but if the radius is a rational multiple of the self-dual
radius, the moduli space contains in addition a certain quotient of
$SU(2)$. On the other hand, for an irrational multiple of the
self-dual radius the additional part of the moduli space is just a
line segment. The bulk operator that changes the radius is exactly
marginal for the bulk theory, but in the presence of certain D-branes
it is not. In particular, it ceases to be exactly marginal if we
consider a rational multiple of the self-dual radius and a D-brane
which is neither Dirichlet or Neumann, but which is associated to a
generic group element $g$ of $SU(2)$. If we change the radius
infinitesimally, it is generically not a rational multiple of
the self-dual radius any more, and thus the brane associated to $g$ is
no longer conformal.  
\smallskip

In order to understand the response of the system to the bulk
perturbation we set up the renormalisation group (RG) equations for
bulk and boundary couplings. This can be done quite generally, and we
find that whenever certain bulk-boundary coupling constants do not
vanish, the exactly marginal bulk perturbation is not exactly marginal
in the presence of a boundary, but rather induces a non-trivial
RG flow on the boundary. In particular, this therefore gives a 
criterion for when an exactly marginal bulk deformation is also
exactly marginal in the presence of a boundary.

For the above example of the free boson, the resulting RG flow
equations can actually be studied in quite some detail. We find that
upon changing the radius the resulting flow drives the brane 
associated to a generic group element $g$ (that only exists at
rational radii) to a superposition of pure Neumann or Dirichlet branes
(that always exist). Whether the end-point is Dirichlet or Neumann
depends on the sign of the perturbation, \ie\ on whether the radius is
increased or decreased. At the self-dual radius, the theory is
equivalent to the $SU(2)$ WZW model at level $1$, and the analysis can
be done very elegantly. In this case we can actually give a closed
formula for the boundary flow which is exact in the
boundary coupling (at first order in the bulk coupling). 

Some of these results can be easily generalised to arbitrary 
current-current deformations of WZW models at higher level and higher 
rank. While we cannot, in general, give an explicit description of the 
whole flow any more, we can still describe at least qualitatively the
end-point of the boundary RG flow.
\medskip

The paper is organised as follows. In section~2 we derive the 
renormalisation group equations that mix bulk and boundary
couplings. In section~3 we apply these techniques to the free boson at
the self-dual radius, and find the exact RG flow. Section~4 discusses
how these results can be generalised to other rational radii, as well
as to current-current deformations of WZW models of higher level and
rank. We conclude in section~5.

\section{The renormalisation group equation}

In this section we shall analyse the RG flow involving bulk and
boundary couplings. Bulk perturbations by relevant operators for
conformal field theories with boundaries have been considered before
in the context of integrable models starting
from~\cite{Cherednik:1985vs} and further developed
in~\cite{Sklyanin:1988yz,Fring:1993mp,Ghoshal:1993tm}. In particular,
these flows have been studied using (an appropriate version of) the
thermodynamic Bethe ansatz (see \eg
\cite{LeClair:1995uf,Lesage,Dorey:1997yg,Dorey:1999cj,Dorey:2004}), in
terms of the truncated conformal space approach (see \eg
\cite{Dorey:1997yg,Dorey:1999cj,Dorey:2000}), and recently by a form
factor expansion~\cite{Bajnok:2006ze,Castro-Alvaredo:2006sh}.
\smallskip

Let $S^\ast$ be the action of a conformal field theory on the upper
half plane. We denote the bulk fields by $\phi_i$, and the boundary
fields by $\psi_j$. Their operator product expansions are of the form 
\begin{eqnarray}
\phi_i(z)\phi_j(w) & = & |z-w|^{h_k-h_i-h_j}\,
C_{ijk}\, \phi_k(w)+\cdots \ ,\\
\psi_i(x)\psi_j(y) & = & (x-y)^{h_k-h_i-h_j}\, 
D_{ijk}\, \psi_k(y)+\cdots \ ,
\end{eqnarray}
where $C_{ijk}$ and $D_{ijk}$ are the bulk and boundary OPE 
coefficients, respectively. (For a general introduction to conformal
field theory see for example \cite{DiFrancesco}.) 
We are interested in the perturbation of
this theory by bulk and boundary fields, 
\begin{equation}
S=S^\ast +\sum_i\tilde\lambda_i\int\phi_i(z)\, d^2z
+\sum_j\tilde\mu_j\int\psi_j(x)\, dx\ . \label{Spert}
\end{equation}
Introducing the length scale $\ell$, we define dimensionless coupling
constants $\lambda_i$ and $\mu_j$ by 
\begin{equation}
\tilde \lambda_i=\lambda_i \, \ell^{h_{\phi_i}-2}\ , 
\qquad 
\tilde\mu_j=\mu_j \, \ell^{h_{\psi_j}-1}\ .
\end{equation} 
Note that we do not assume here that $\phi_i$ and $\psi_j$ are
marginal operators. If we expand the free energy in powers of  
$\lambda_i$ and $\mu_j$, we get terms of the form
\begin{multline}
\frac{\lambda_1^{l_1}\cdots\mu_1^{m_1}\cdots}{l_1!\cdots m_1!\cdots}
\prod_i \ell^{(h_{\phi_i}-2)l_i}\prod_j \ell^{(h_{\psi_j}-1)m_j} \\
\times\int\langle\phi_1(z_1^1) \phi_1(z_2^1)\cdots\phi_2(z^2_1) \cdots
\psi_1(x_1^1)\cdots\rangle \prod d^2z_k^i \prod dx_k^j \ .\label{int}
\end{multline}
To regularise (\ref{int}), we use an UV cutoff $\ell$. More
precisely, the prescription is
\be\label{UV}
|z^i_k-z^{i'}_{k'}| > \ell \ , \qquad 
|x^j_k-x^{j'}_{k'}| > \ell \ , \qquad \Im z > \frac{\ell}{2} \ . 
\ee
The parameter $\ell$ thus appears in (\ref{int}) both explicitly as
powers in $h$, and implicitly through the range of integration.  

Following \cite{Cardy} we now consider a change of the scale $\ell$,
$\ell \rightarrow (1+\delta t)\ell$, and ask how the coupling constants have
to be adjusted so as to leave the free energy unchanged.  The explicit
dependence of the expression~\eqref{int} on $\ell$ leads to a change in
$\lambda_{i}$ and $\mu_{j}$ by
\begin{eqnarray}
\lambda_i \rightarrow (1+(2-h_{\phi_i})\, \delta t)\, \lambda_i\ ,\nn\\
\mu_j \rightarrow (1+(1-h_{\psi_j})\, \delta t)\, \mu_j\ . 
\end{eqnarray}
The implicit dependence of~\eqref{int} on $\ell$ through the UV
prescription (\ref{UV}) gives rise to an additional change of the
coupling constants. From the first inequality in (\ref{UV}), which
controls the UV singularity in the bulk operator product expansion, we
obtain the equation 
$\delta\lambda_k = \pi C_{ijk}\lambda_i\lambda_j \delta t$ 
\cite{Cardy}. A similar calculation gives 
$\delta\mu_k = D_{ijk}\mu_i\mu_j \delta t$ (see for example
\cite{Affleck:1991tk}) for the  
contribution from the boundary operator product expansion (the second
inequality). Finally we have to consider the contribution from the
third inequality which controls the singularity that arises when a 
bulk operator approaches the boundary. When we scale $\ell$ by 
$(1+\delta t)$ we change the integration region of a bulk operator by
a strip parallel to the real axis of width $\ell\, \delta t/2$. This
changes the expression~\eqref{int} by terms of the form  
\begin{equation}\label{term}
- \lambda_i \, \ell^{h_{\phi_i}-2}\int dx \, 
\int_{\ell/2}^{\ell/2 + \ell \delta t/2} dy \, 
\langle\cdots\phi_i(z)\cdots\rangle \ ,
\end{equation}
where we have written $z=x+iy$. In order to evaluate this
contribution, we use the bulk-boundary operator product expansion   
\begin{equation}
\phi_i(z,\bar z) = (2y)^{h_{\psi_j}-h_{\phi_i}}\, B_{ij}
\, \psi_j(x) + \cdots  \ , 
\end{equation}
where $B_{ij}$ is the bulk-boundary OPE coefficient that depends on
the boundary condition in question. The change of the free energy
described by (\ref{term}) is then
\begin{equation}
-\lambda_i \, \ell^{h_{\phi_i}-2}\int dx\, 
\frac{\ell \, \delta t}{2}\, B_{ij} \, \ell^{h_{\psi_j}-h_{\phi_i}} \, 
\langle\cdots\psi_j(x)\cdots\rangle = - \frac{1}{2} \, B_{ij} \, 
\ell^{h_{\psi_j}-1}\, \lambda_i \, \delta t \int dx\, 
\langle\cdots\psi_j(x)\cdots\rangle \label{B}
\end{equation}
which can be absorbed by a shift of 
$\delta\mu_j=\tfrac{1}{2}\, \lambda_i B_{ij} \delta t$.
Collecting all terms, we thus obtain the RG equations to
lowest order 
\begin{eqnarray}
\label{RGbulk}
\dot\lambda_k &=& (2-h_{\phi_k})\lambda_k +
\pi C_{ijk}\, \lambda_i\lambda_j + {\cal O}(\lambda^3)\ , \\
\dot\mu_k &=& (1-h_{\psi_k})\mu_k + \frac{1}{2}\, B_{ik}\, \lambda_i  
+ D_{ijk}\, \mu_i\mu_j + {\cal O}(\mu\lambda, \mu^3, \lambda^2)\ .
\label{bounflow}
\end{eqnarray}
The flow of the bulk variables $\lambda_{k}$ in~\eqref{RGbulk} is
independent of the boundary couplings $\mu_{k}$ on the disc. The RG
flow in the bulk therefore does not depend on the boundary condition
whereas the bulk has significant influence on the flow of the boundary
couplings.  Note that the terms we have written out explicitly are
independent of the precise details of the UV cutoff (if the fields are
marginal). Higher order corrections, on the other hand, will depend on
the specific regularisation scheme.
\medskip

Suppose now that $\phi_i$ is an exactly marginal bulk
perturbation. The perturbation by $\phi_i$ is then exactly marginal in
the presence of a boundary if the bulk boundary coupling constants 
$B_{ik}$ vanish; this has to be the case for all boundary fields
$\psi_k$ (except for the vacuum) that are relevant or marginal, 
{\it i.e.} satisfy $h_{\psi_k}\leq 1$. Obviously, switching on
the vacuum on the boundary just leads to a rescaling of the disc
amplitude; for irrelevant operators, on the other hand, the flow is 
damped by the first term of (\ref{bounflow}), and thus the bulk
perturbation only leads to a small correction of the boundary
condition. 

The above condition is the analogue of the usual statement about exact
marginality: 
a necessary condition for a marginal bulk (boundary) operator to be
exactly marginal is that the three point couplings $C_{iik}$
($D_{iik}$) vanish for all marginal or relevant fields $\phi_k$
($\psi_k$), except for the identity (see for example 
\cite{Kadanoff:1978pv,Cardy,Recknagel:1998ih}). 

If the bulk boundary coefficient $B_{ik}$ does not vanish for some
relevant or marginal boundary operator $\psi_{k}$, the corresponding
boundary coupling $\mu_{k}$ starts to run, and there is a non-trivial
RG flow on the boundary. The bulk couplings $\lambda_{i}$ are not
affected by the flow ($\dot{\lambda}_{i}=0$), and we can thus
interpret it as a pure boundary flow in the marginally deformed bulk
model. From that point of view it is then clear that the flow must
respect the g-theorem~\cite{Affleck:1991tk,Friedan:2003yc}. In
particular, the g-function of the resulting brane is smaller than that
of the initial brane. This is in fact readily verified for the examples
we are about to study.

\section{The free boson theory at the self-dual radius}
\setcounter{equation}{0}

As an application of these ideas, we now consider the example of the 
free boson theory at $c=1$. We shall first consider the theory at the
critical radius, where it is in fact equivalent to the WZW model of
$su(2)$ at level $1$. For this theory all conformal boundary states
are known~\cite{Gaberdiel:2001xm}, and are labelled by group elements
$g\in SU(2)$ (for earlier work see  also
\cite{Callan:1994ub,Polchinski:1994my}). 

\begin{figure}
\begin{center}
\epsfig{file=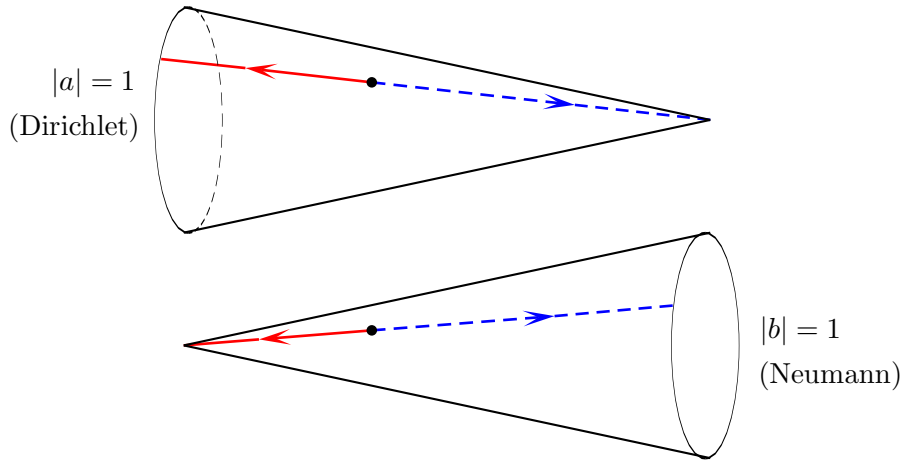}
\end{center}
\caption{\label{schultuete}The moduli space of D-branes on the
self-dual circle, $SU(2)$, can be described as a product of two
circles $S^{1}$ (given by the phases of $a$ and $b$ in~\eqref{su2})
fibred over an interval where $|a|$ runs between $0$ and $1$, and
$|a|^{2}+|b|^{2}=1$. The ends of the interval where one of the circles
shrinks to zero describe Dirichlet and Neumann branes, respectively.
If we start with a generic boundary condition and increase (decrease)
the radius, the boundary condition will flow to a Dirichlet (Neumann)
boundary condition.}
\end{figure}

Suppose that we are considering the boundary condition labelled by
$g\in SU(2)$, where we write 
\begin{equation}\label{su2}
g = \left( \begin{matrix}
a & b^\ast \cr - b & a^\ast 
\end{matrix} \right) \ , 
\end{equation}
and $a$ and $b$ are complex numbers satisfying $|a|^2+|b|^2=1$.
(Geometrically, $SU(2)$ can be thought of as a product of two circles
--- see figure~\ref{schultuete}.) We shall choose the convention that
the brane labelled by $g$ satisfies the gluing condition\footnote{Note
that the labelling differs from the one used
in~\cite{Gaberdiel:2001zq}.}
\begin{equation}\label{gluing}
\left( g\, J^{\alpha}_m \, g^{-1} + \bar{J}^{\alpha}_{-m} \right) \, 
|\!| g \rangle\!\rangle = 0 \ , 
\end{equation}
where $J^{\alpha}$ are the currents of the WZW model (the
corresponding Lie algebra generators will be denoted by $t^{\alpha}$).
We shall furthermore use the identification that $g$ diagonal ($b=0$)
describes a Dirichlet brane on the circle, whose position is given by
the phase of $a$; conversely, if $g$ is off-diagonal ($a=0$), the
brane is a Neumann brane, whose Wilson line on the dual circle is
described by the phase of $b$.

\subsection{Changing the radius}

\noindent We want to consider the bulk perturbation by the field
\begin{equation}
\Phi = J^3 \bar{J}^3 \ , \qquad \hbox{where} \qquad 
t^3 = \frac{1}{\sqrt{2}} \, 
\left( \begin{matrix} 1 & 0 \cr 0 & -1 \end{matrix} \right) \ . 
\end{equation}
This is an exactly marginal bulk perturbation that changes the radius
of the underlying circle. With the above conventions, the perturbation
$\lambda \Phi$ with $\lambda>0$ increases the radius, while
$\lambda<0$ decreases it. At any rate, the perturbation by $\Phi$
breaks the $su(2)$ symmetry down to $u(1)$. However, in the presence
of a boundary, the bulk perturbation is generically not exactly
marginal any more. This is implicit in the results of
\cite{Gaberdiel:2001zq,Janik,Friedan} since the set of possible
conformal boundary conditions is much smaller at generic (irrational)
radius relative to the self-dual case. Here we want to study in detail
what happens to a generic boundary condition under this bulk
deformation.   

Even before studying the detailed RG equations that we derived in the
previous section, it is not difficult to see that the above
deformation is generically not exactly marginal. In
particular, we can consider the perturbed one-point function of the
field $\Phi$ in the presence of the boundary. To first order, this
means evaluating the 2-point function
\begin{equation}
\lambda\int_{\mathbb{H}^+}d^2z 
\langle (J^{\alpha}\bar J^{\alpha})(z)\, 
(J^{\alpha}\bar J^{\alpha})(w)\rangle \label{1pt1l} \ , 
\end{equation}
where the label $\alpha=3$ is not summed over. Using the usual
doubling trick \cite{Cardy:1984bb} this amplitude can be expressed as
a chiral 4-point function, where we have the fields $J^{\alpha}$ at
$z$ and $w$, and the `reflected' fields 
$J^{\beta}\equiv g J^{\alpha} g^{-1}$ at $\bar{z}$ and $\bar{w}$. 

\noindent The chiral correlation functions of WZW models at level $k$
can be calculated using the techniques of
\cite{FZ,Gaberdiel:1998fs}. 
Let $t^{\alpha}, \ \alpha=1,\ldots, {\rm dim} (g)$, be the Lie algebra
generator (corresponding to $J^{\alpha}$) in some representation; we
choose the normalisation
\begin{equation}
\Tr(t^{\alpha}\, t^{\beta})=k\, \delta^{\alpha \beta}\ .
\end{equation}
To evaluate 
$\langle J^{\alpha_1}(z_1)\cdots J^{\alpha_n}(z_n) \rangle$, consider 
then all permutations $\rho \in S_n$ that have no fixed points; this
subset of permutations is denoted by $\tilde{S}_n$. Each such $\rho$
can be written as a product of disjoint cycles
\begin{equation}
\rho = \sigma_1 \sigma_2 \cdots \sigma_M \ .
\end{equation}
To each cycle $\sigma = (i_1\,i_2\cdots i_m)$ we assign the function 
\begin{equation}
f_{\sigma}^{\,\alpha_{i_1}\cdots \alpha_{i_m}}(z_{i_1}, \ldots, z_{i_m})
= -\frac{\Tr( t^{\,\alpha_{i_1}}\cdots t^{\,\alpha_{i_m}})}
{(z_{i_1}-z_{i_2})(z_{i_2}-z_{i_3}) \cdots (z_{i_m}-z_{i_1})} \ ,
\end{equation}
and to each $\rho$ the product $f_{\sigma_1}\cdots f_{\sigma_M}$.
The correlation function is then given by summing over all permutations
without fixed points, 
\begin{equation}\label{correlators}
\langle J^{\alpha_1}(z_1)\cdots J^{\alpha_n}(z_n) \rangle = 
\sum_{\rho \in \tilde S_n} f_\rho \ .
\end{equation}

\noindent In (\ref{1pt1l}), $\rho$ is either a 4-cycle or consists
of two 2-cycles. In the latter case we get the terms 
\begin{equation}
\frac{(\Tr(t^{\alpha} t^{\beta}))^2}{|z-\bar z|^2|w-\bar w|^2}+
\frac{(\Tr(t^{\alpha} t^{\beta}))^2}{|z-\bar w|^4} +
\frac{\Tr(t^{\alpha} t^{\alpha})\Tr(t^{\beta} t^{\beta})}{|z- w|^4} \ .
\end{equation}
Integration over the upper half plane gives (divergent) contributions
proportional to 
\makebox{$|w-\bar w|^{-2}$,} which can be absorbed in the
renormalisation of $J^{\alpha}$. 
The six terms that come from the six different 4-cycles 
give a total contribution of  
\begin{equation}
-\frac{\Tr([t^{\alpha},t^{\beta}]^2)}
{(z-\bar z)(w-\bar w)| z-\bar w|^2} \ .
\end{equation}
Set $w=i|w|$ and $z=x+iy$. The resulting integral over the upper half
plane is logarithmically divergent for $y\rightarrow 0$. Introducing
an ultraviolet cutoff $\epsilon$, we get   
\begin{equation}
\int_\mathbb{R}dx\int_\epsilon^\infty\!\!\! dy\, \frac{1}{2iy2i|w|}
\frac{1}{x^2+(y+|w|)^2} 
= \frac{\pi}{4|w|^2}\log \epsilon
      -\frac{\pi}{8|w|^2}\log |w|^2 +\cal{O}(\epsilon)\ .
\end{equation}
The first term has the right $w$ dependence to be absorbed by a
suitable renormalisation of $J^{\alpha}$. The second term, however,
pushes the conformal weight away from $(1,1)$.  Thus, if $J^{\alpha}$
is to be exactly marginal, the expression
$\Tr([t^{\alpha},t^{\beta}]^2)$ must vanish.  
\smallskip

\noindent In the case above $\Tr([t^{\alpha},t^{\beta}]^2)$ equals
\begin{equation}
\Tr([t^3,g\, t^3\, g^{-1}]^2)=- 8 |a|^2|b|^2 \ .
\end{equation}
This only vanishes if either $|a|=0$ or $|b|=0$; the corresponding
boundary conditions are therefore either pure Dirichlet or pure 
Neumann boundary conditions. This ties in with the expectations based
on the analysis of the conformal boundary conditions since 
only pure Neumann or Dirichlet boundary conditions exist for all
values of the radius.
\smallskip

The argument above can also be used in the general case to derive a
necessary criterion for when a bulk deformation is exactly marginal in
the presence of a boundary. It is not difficult to see that it leads
to the same criterion as the one given in section~2.

\subsection{The renormalisation group analysis}

Now we want to analyse what happens if $g$ does not describe a pure
Neumann or pure Dirichlet boundary condition. In particular, we can
use the results of section~2 to understand how the system
reacts to the bulk perturbation by $\lambda \Phi$.

In order to see how the boundary theory is affected by the
perturbation we have to compute the bulk boundary OPE of the
perturbing field $\Phi$. There are no relevant boundary fields (except
the vacuum), and the marginal fields are all given by boundary
currents $J^{\gamma}$. We can thus determine the bulk boundary 
OPE coefficient $B_{\Phi \gamma}$ from the two-point function
\begin{equation}
\langle J^{\gamma} (x) (J^{3}\bar{J}^{3}) (z) \rangle = 
B_{\Phi \gamma} \, |z-\bar{z}|^{-1} |x-z|^{-2}  \ ,
\end{equation}
which -- employing the general formula~\eqref{correlators} -- leads to
\begin{equation}
B_{\Phi \gamma} = -i \Tr (t^{\gamma}[t^{3},g\,  t^{3}g^{-1} ])\ .
\end{equation}
We see that the only boundary field that is switched on by the bulk
perturbation is the current $J^{\gamma}$ whose (hermitian) Lie algebra
generator $t^{\gamma}$ is proportional to the
commutator~$[t^{3},g\,t^{3}g^{-1}]$. The normalised $t^{\gamma}$ is
given by
\begin{equation}
t^{\gamma} = \frac{i}{\sqrt{2}} \left(
\begin{matrix}
0 & - e^{i\chi} \cr
e^{-i\chi} & 0 
\end{matrix} \right) \qquad \hbox{with} \qquad
a \, b^\ast = |a b| e^{i\chi} \ . 
\end{equation}
Its relation to the commutator is  
\begin{equation}
- i [t^3, g \, t^3 g^{-1} ] = 
- i \left( \begin{matrix}
0 & - 2 a b^\ast \cr  2 a^\ast b & 0 \end{matrix}
\right)  = B \, t^{\gamma} \ ,
\end{equation}
where the bulk boundary coefficient $B=B_{\Phi \gamma}$ is given by
\begin{equation}
B = - 2 \, \sqrt{2}\, |a| \, |b| \ .
\end{equation}
The boundary current proportional to $t^{\gamma}$ modifies the boundary  
condition $g$ by  
\begin{equation} \label{deltag}
\delta g = i\, t^{\gamma} \, g =  
 \frac{1}{\sqrt{2}}
\left(
\begin{matrix}
- a \frac{|b|}{|a|} &    b^\ast \frac{|a|}{|b|} 
\vspace*{0.2cm} \cr
- b \frac{|a|}{|b|}   &  - a^\ast \frac{|b|}{|a|} 
\end{matrix} \right) \ . 
\end{equation}
This leaves the phases of $a$ and $b$ unmodified, but decreases the
modulus of $a$ while increasing that of $b$. 

Since the operators are marginal, the renormalisation group 
equation to lowest order in the coupling constants (\ref{bounflow})
is now 
\begin{equation}\label{linear}
\dot{\mu} = \frac{1}{2}\, B\, \lambda 
+ {\cal O} (\mu\lambda,\mu^2,\lambda^2)  \ ,
\end{equation}
where $\mu$ is the boundary coupling constant of the field
$J^{\gamma}$.  Thus if the radius is increased ($\lambda >0$), $\mu$
becomes negative, and the boundary condition flows to the boundary
condition with $b=0$ --- the resulting brane is then a Dirichlet brane
whose position is determined by the original phase of $a$. Conversely,
if the radius is decreased ($\lambda <0$), $\mu$ becomes positive, and
the boundary condition flows to the boundary condition with $a=0$. The
resulting brane is then a Neumann brane whose Wilson line is
determined by the original value of the phase of $b$ (see
figure~\ref{schultuete}). This is precisely what one should have
expected since for radii larger than the self-dual radius, only the
Dirichlet branes are stable, while for radii less then the self-dual
radius, only Neumann branes are stable.  
\medskip

\noindent Actually, the renormalisation group flow can be studied in
more detail. It follows from (\ref{deltag}) that to lowest order in
$\mu$   
\begin{equation}\label{amu}
a(\mu) = a_0 - \mu \, a_0 \frac{|b_0|}{\sqrt{2} |a_0|} 
+ {\cal O}(\mu^2) \ , 
\end{equation}
where the initial values of $a$ and $b$ have been denoted by $a_0$ and
$b_0$, respectively. Since $a$ depends on the RG parameter only via
$\mu$, it thus follows that 
\begin{equation}
\dot{a} = - \dot{\mu} \, a \, \frac{|b|}{\sqrt{2} |a|} = -
\frac{B}{2\sqrt{2}} \, \frac{|b|}{|a|} \, a \, \lambda 
= |b|^2 \, a \, \lambda = (1 - |a|^2) \, a \, \lambda \ . \label{mudot}
\end{equation}
If we write $|a|=\sin \psi$, this simplifies to 
\begin{equation}
\dot{\psi} = \sin \psi \, \cos \psi \, \lambda \ . 
\end{equation}
Denoting the RG parameter by $t$, the solution to this differential
equation is 
\begin{equation}
\tan \psi (t) = \tan \psi (0) \, e^{\lambda t} \ .
\end{equation}
Thus for $\lambda >0$ this flows indeed to $|a_{\infty}|=1$, while for
$\lambda <0$ we find $|a_{\infty}|=0$, as expected.

Given the relation (\ref{amu}), we can deduce from the solution 
for $a(t)$ a differential equation for $\mu(t)$ which turns out to be
\begin{equation}
\dot \mu = -\sqrt{2} \, \dot{\psi} \ .
\end{equation}
This can be integrated to 
\begin{equation}\label{muoft}
\mu(t)=-\sqrt{2}\,( \psi (t) -\psi (0))\ . 
\end{equation}
We can thus determine the path on the group manifold as 
\begin{equation}
g(t) = e^{i \mu(t) \, t^{\gamma}} \, g \ .
\end{equation}
As a consistency check one verifies that 
\begin{equation}
\lim_{t\rightarrow \infty} g(t) = 
\left\{
\begin{array}{cl}
 \left( \begin{array}{cc}   \frac{a}{|a|} & 0 \\
0 & \frac{a^\ast}{|a|} \end{array} \right )
 & \qquad \hbox{if $\lambda >0$}\vspace*{0.2cm} \\
\left( \begin{array}{cc} 0 & \frac{b^\ast}{|b|} \\
-\frac{b}{|b|} &0 \end{array} \right )
 & \qquad \hbox{if $\lambda <0$.}
\end{array}
\right. 
\end{equation}
The path is actually a geodesic on $SU(2)$, relating the point $g$ to
the nearest diagonal or off-diagonal group element. In order to see
this we write 
\begin{equation}
g = \left(
\begin{matrix}
\sin \psi \, e^{i\theta} &
\cos \psi \,  e^{-i\varphi} \\
-\cos \psi \, e^{i\varphi} & 
\sin \psi e^{-i\theta}
\end{matrix} \right) \ ,  
\end{equation}
where $0\leq \psi \leq \tfrac{\pi}{2}$ and 
$0\leq \theta, \varphi < 2 \pi$. In these
variables, the metric on $SU(2)$ is 
\begin{equation}
ds^2 = d\psi^2 + \sin^2 \psi \,  d\theta^2 + 
\cos^2\psi \, d\varphi^2 \ .
\end{equation}
The above path in $SU(2)$ is the path with $\theta$ and $\varphi$
constant. The variable $\mu$ (see eq.\ \eqref{muoft}) is simply
proportional to $\psi - \psi_0$, which is the arc length parameter
along the curve.

\section{Generalisations}
\setcounter{equation}{0}

It is not difficult to generalise the above analysis in a number of
different ways.

\subsection{The free boson away from criticality}

If the radius of the free boson is a rational multiple of the
self-dual radius, $R=\tfrac{M}{N} R_{\rm sd}$, then a similar analysis
applies. At this radius, the conformal boundary states are labelled by
elements in the quotient space
\begin{equation}
g \in  SU(2) / \Zop_M \times \Zop_N \ , 
\end{equation}
where $\Zop_M$ and $\Zop_N$ act by multiplication by roots of unity on
$a$ and $b$, respectively, leaving the absolute values unaffected
\cite{Gaberdiel:2001zq} 
One way to arrive at this construction is to describe the
theory at radius $R$ as a freely-acting orbifold by 
$\Zop_M\times \Zop_N$ of the self-dual radius theory
\cite{Tseng:2002ax}. Under this orbifold action
none of the generic $SU(2)$ branes are invariant, and thus the branes
of the orbifold are simply the superpositions of $MN$ branes of the
$SU(2)$ level $1$ theory.

In particular, it therefore follows that the bulk-boundary 
OPE coefficients that were relevant in the above
analysis are (up to an $MN$ dependent factor) unmodified. Therefore the same
conclusions as above hold: if the radius is increased, 
a generic brane flows to $M$ equally spaced Dirichlet branes 
(this is the interpretation of the branes with $b=0$); if the radius
is decreased, a generic brane flows to $N$ Neumann branes whose Wilson
lines are equally spaced on the dual circle ({\it i.e.} the branes
with $a=0$). Since the phases of $a$ and $b$ are unchanged along the
flow, the flow is obviously compatible with the $\Zop_M\times \Zop_N$
orbifold operation that only acts on these phases.

\subsection{The analysis at higher level}

For $SU(2)$ at level $k$, the branes that preserve the affine symmetry
(up to an inner automorphism by conjugation by a group element $g\in
SU(2)$) are labelled by $|\!| j , g \rangle\!\rangle$, where
$j=0,\tfrac{1}{2},1,\ldots, \tfrac{k}{2}$ denotes the different
representations of $\hat{su}(2)$ at level $k$ (that label the
different Cardy branes~\cite{Cardy:1989ir}), while $g$ describes the
automorphism
\begin{equation}
\left(g \, J^{\alpha}_m \, g^{-1} + \bar{J}^{\alpha}_{-m} \right)
\, |\!| j , g \rangle\!\rangle  = 0 \ .
\end{equation}
In addition there is the identification,
\begin{equation}\label{ident}
|\!| j , g \rangle\!\rangle = 
|\!| \tfrac{k}{2} - j, -g \rangle\!\rangle \ , 
\end{equation}
where $-g \in SU(2)$ is minus the $2\times 2$ matrix 
(\ref{su2}).  

The field $\Phi$ is an exactly marginal bulk field for any level $k$
\cite{Chaudhuri:1988qb,Hassan:1992gi}. We can thus ask what happens to
the boundary condition $|\!| j , g \rangle\!\rangle$ as we perturb the
theory by $\Phi$. 

In fact, it is easy to see that the above analysis for level 1 still
goes through --- the only place where $k$ enters is in the overall
normalisation of the bulk-boundary OPE coefficient that is largely
irrelevant for our analysis. Thus if we perturb the theory by the
exactly marginal bulk perturbation $J^3\bar{J}^3$, the brane labelled
by $|\!| j , g \rangle\!\rangle$ flows to $|\!| j , g_0
\rangle\!\rangle$, where $g_0$ is either diagonal or off-diagonal
(depending on the sign of the bulk coupling constant $\lambda$), and
the relevant phase of $a_0$ or $b_0$ agrees with the original phase of
$a$ or $b$ in $g$, respectively. In particular, this prescription
therefore respects the identification (\ref{ident}). It is also worth 
noting that it does not mix different $j$, and therefore does not
produce any additional flows that would reduce the K-theoretic charge
group \cite{AS,Fredenhagen:2000ei}.

The bulk perturbation breaks the $SU(2)$ symmetry down to $SU(2)/U(1)
\times U(1)$, where the radius of the $U(1)$ factor is deformed away
from the original value of $\sqrt{k}$ times the self-dual radius. The
branes corresponding to $g_0$ (to which any brane will flow) describe
factorisable boundary conditions that define a standard Dirichlet or
Neumann boundary condition for the $U(1)$ factor. It is then clear
that these branes exist for arbitrary radius of this $U(1)$ (this has
been analysed previously in~\cite{Forste:2001gn,Forste:2003ne}). The
resulting picture is therefore again in agreement with expectations.

For large values of the level $k$ we can give yet another geometric
interpretation. The current-current deformation of the WZW model can
be understood as deforming the metric, the B-field and the dilaton on
the group. In particular, once the WZW model is deformed 
the dilaton~$\phi$ is not constant any more,
but has the dependence (see~\cite{Hassan:1992gi,Giveon:1993ph,Forste:2001gn})
\begin{equation}\label{dilaton}
e^{-2\phi (\psi ) } = \frac{1- (1-R^{2})\cos^{2} \psi}{R} \ ,
\end{equation}
where $R$ denotes the deformed radius of the embedded $U(1)$ ($R=1$
being the WZW case). If we start with a D0-brane on the group at
position $g$, then after the deformation it will flow along the gradient of
the dilaton to a maximum, such that its mass, which is proportional to  
$\frac{1}{g_{s}}\sim e^{-\phi }$, is minimal. Minimisation
of~\eqref{dilaton} leads to the conditions 
\begin{equation}
(1-R^{2})\sin 2\psi = 0 \ , \qquad (1-R^{2}) \cos 2\psi >0 \ .
\end{equation}
When the radius is increased ($R>1$, corresponding to $\lambda >0$),
we find $\psi =\frac{\pi}{2}$, \ie\ $|a|=1$. For $R<1$ we obtain 
on the other hand $\psi =0$ ($|b|=1$). This is thus in nice agreement
with our analysis of section~3.

\subsection{Other bulk perturbations}

So far we have only considered bulk perturbations by $J^3 \bar{J}^3$,
but it should be clear how to generalise this to the case where the
perturbing bulk field is $J^{\alpha} \bar{J}^{\bar{\alpha}}$. In fact,
if we write $t^{\alpha} =h t^3 h^{-1}$ and $t^{\bar{\alpha}} = \bar{h}
t^3 \bar{h}^{-1}$, then the above analysis goes through provided we
replace $g$ by $\hat{g} = h^{-1} \, g \, \bar{h}$. Indeed, the
relevant $t^{\gamma}$ is in this case
\begin{equation}
it^{\gamma} \propto  [t^{\alpha}, g \, t^{\bar{\alpha}} \, g^{-1} ] =  
h [ t^3 , \hat{g} \, t^3 \, \hat{g}^{-1} ]  h^{-1} \ , 
\end{equation}
and thus 
\begin{equation}
\delta g= \delta (h \, \hat{g} \, \bar{h}^{-1}) = 
h\,  \delta \hat{g} \, \bar{h}^{-1}  \ .
\end{equation}
At level $1$, the perturbation by 
$J^{\alpha} \bar{J}^{\bar{\alpha}}$ can again be
interpreted as changing the radius of a circle. Its embedding in $SU(2)$   
is described as 
\begin{equation}
\theta \mapsto h e^{i\theta t^{3}} \bar{h}^{-1} \ .
\end{equation}

\subsection{Higher rank groups}

Much of the discussion for $SU(2)$ carries over to Lie groups of
higher rank, though in general it is not possible to give a closed
expression for the integrated flow any more. For simplicity we shall
restrict the following discussion to the Lie groups $G = SU(n)$.  

Let us consider a D-brane that is characterised by the gluing
condition~\eqref{gluing} for a given $g\in SU (n)$. As in
section~3.1, the perturbation $J^{\alpha}\bar{J}^{\alpha}$ with 
$\alpha$ fixed and $t^{\alpha} \in su(n)$ is exactly
marginal in the bulk \cite{Chaudhuri:1988qb,Hassan:1992gi}, but leads
to a flow of the gluing parameter $g$ as
\begin{equation}\label{gdot}
\dot g = \frac{\lambda}{2}\, [t^{\alpha},t^{\beta}]\, g\ , 
\end{equation}
where $t^{\beta}=g\,t^{\alpha}\,g^{-1}$. This flow can be interpreted
as a gradient flow,
\begin{equation}\label{gradientflow}
\dot{g} = - \nabla V (g)\qquad \hbox{with potential} \qquad
V (g) = -\frac{\lambda}{2} \Tr (t^{\alpha}\,g \,t^{\alpha}\, g^{-1})\ .
\end{equation}
To see this, we first recall that the gradient is defined by  
\begin{equation}\label{gradient1}
\left.\frac{d}{ds} V(g+istg) \right|_{s=0} = 
- \Tr (\nabla V(g) \, g^{-1} \, i t) \ , 
\end{equation}
where $t$ is an arbitrary vector in the Lie algebra. Here the minus
sign appears because the trace is negative definite on the Lie
algebra; the factors of $g$ map $it$ to a tangent vector $itg$ at
$g$, and the tangent vector $\nabla V(g)$ to an element of the Lie
algebra, $\nabla V(g) \, g^{-1}$. Evaluating the directional
derivative we find
\begin{align}
\left.\frac{d}{ds} V(g+istg) \right|_{s=0} 
&= -\frac{\lambda}{2} 
\Tr \big( 
t^{\alpha}\, itg \, t^{\alpha}g^{-1}
-t^{\alpha} gt^{\alpha}\, g^{-1}it\big) 
\nonumber \\
&              = \frac{\lambda}{2} 
\Tr \big( [t^{\alpha},gt^{\alpha}g^{-1}]\, g \, (g^{-1}it) \big) \ . 
\end{align}
Comparing this with~\eqref{gradient1} we deduce that 
\begin{equation}
\nabla V (g) = - \frac{\lambda}{2} \, [t^\alpha,g\, t^\alpha\,g^{-1}]
\, g \ ,
\end{equation}
which hence implies that \eqref{gradientflow} reproduces the flow
equation~\eqref{gdot}.  

In contradistinction to the $SU(2)$ case, however, this flow
is generically not a geodesic flow. The change of the direction of the
RG flow is 
\begin{equation}
\frac{d}{dt} [t^{\alpha},t^{\beta}] \propto
[t^{\alpha},[t^{\beta},[t^{\alpha},t^{\beta}]]]
\end{equation}
which is in general not proportional to $[t^{\alpha},t^{\beta}]$. Thus
the tangent to the flow is not parallel to a fixed direction in the
Lie algebra; this makes it hard to integrate the complete flow in the
generic case.  

We can nevertheless describe at least qualitatively the
end point of the flow. To this end it is sufficient to understand
the fixed points of the flow and their stability
properties. 

A boundary condition corresponding to the gluing condition $g$ is a
fixed point of the flow if $[t^{\alpha}, t^{\beta}]=0$. This is only
the case if the matrices $t^{\alpha}$ and $t^{\beta}$ have common
eigenspaces. Assume that $t^{\alpha}$ is generic, \ie that all its
eigenvalues $\tau_{i}$ are distinct and all eigenspaces
$\mathbb{R}v_{i}$ are one-dimensional. Then 
$[t^{\alpha}, t^{\beta}]=0$ if and only if $g$ permutes the $n$
eigenspaces and 
multiplies each one by a phase $r_i$. This means that there are $n!$
discrete choices for $g$, each coming with $n-1$ continuous degrees of
freedom (note that $\det g= \pm \prod_{i} r_{i}=1$).

This has a simple physical interpretation if the level of the WZW
model is $1$. Then the theory is equivalent to a compactification on a
torus described by the momentum lattice
\begin{equation}
\{(p_{L},p_{R})\in \Lambda_{W}\oplus \Lambda_{W}\ ,
\ p_{L}-p_{R}\in \Lambda_{R}\} \ ,
\end{equation}
where $\Lambda_{W}$ and $\Lambda_{R}$ are the weight and root lattice
of $su (n)$, respectively. Without loss of generality we may choose
our Cartan subalgebra such that it contains $t^{\alpha}$.
A group element $g\in SU (n)$ that permutes the eigenvectors $v_{i}$
acts by conjugation on the root lattice and hence corresponds to some
element $w_{g}$ of the Weyl group. The gluing condition~\eqref{gluing}
for the currents $J^{\beta}$ then translates into the condition
\begin{equation}
w_{g} p_{L} = p_{R} 
\end{equation}
for the momenta. This is the gluing condition for the standard torus 
branes that couple to all momenta $p_{L}$ 
(as $w_{g}p_{L}-p_{L}\in \Lambda_{R}$).
The dimension of the brane is given by the 
number of eigenvalues of $w_{g}$ that are not equal to $1$ (this is
the absolute length of $w_{g}$). The phases of $g$ then correspond to
the positions and Wilson lines of the brane.

These standard torus D-branes are the ones that are unaffected by a
perturbation of the size of the torus and they correspond to the fixed
points $g$ of the flow equation~\eqref{gdot}.
\medskip

In order to understand where a generic brane flows to, it is
furthermore important to understand the stability of the fixed
points. 
Suppose we start with a boundary condition that is very close to one
of the fixed points; if the brane is driven back to the 
fixed point it is {\it stable}, if it flows away (to some other
fixed point) it is {\it unstable}.
 
To simplify the discussion we shall work in the 
eigenbasis $\{v_{i}\}$ of $t^{\alpha}$. Using its spectral decomposition  
$t^{\alpha}=\sum \tau_i\, P_i$, we can rewrite~(\ref{gdot}) as
\begin{equation} 
\dot g = \frac{\lambda}{2}\sum_{i, j}\tau_i \tau_j (P_i\,g\,
P_j -g\, P_i\, g^{-1} P_j \, g)\ . \label{gdoteig}
\end{equation}
To check the stability of a fixed point $g=S$, consider the ansatz
\begin{equation}
g_{ij}(t)= S_{ij}+\epsilon\, h_{ij}(t)\ .
\end{equation}
Here $S$ is the matrix of a fixed point given by a permutation
$\sigma$ and phases $r_{i}$, \ie
\begin{equation}
S:\, v_i\mapsto r_i\, v_{\sigma(i)} \ .
\end{equation}
In particular, this means that 
\begin{equation}
 S\, P_i \,S^{-1} = P_{\sigma(i)}\ .
\end{equation}
Evaluating (\ref{gdoteig}) to first order yields
\begin{equation} \label{hstable}
\dot h_{ij} = \frac{\lambda}{2}\,
(\tau_i -\tau_{\sigma(j)})\, 
(\tau_j-\tau_{\sigma^{-1}(i)})\, h_{ij}\ .
\end{equation}
We easily see that $\dot{h}_{ij}=0$ for $i=\sigma (j)$; these are the 
$n-1$ flat directions we have identified before. In order for $g=S$ to
be stable, all other components $h_{lm}$ must have negative
eigenvalues. Without loss of generality, we may assume that 
the $\tau_i$ are ordered,
\begin{equation}
\tau_1 < \tau_2 < \cdots < \tau_n\ .
\end{equation}
Consider then the coefficient for $i=\sigma(p)$. If $\lambda > 0$ the 
condition is
\begin{equation}
j < p \Rightarrow \sigma(j) < \sigma(p)\ ,
\end{equation}
\ie $\sigma$ grows monotonically, which is only the case for 
$\sigma ={\rm id}$. For $\lambda <0$, $\sigma$ must be a decreasing
function, 
\ie
\begin{equation}
\sigma\, : i \mapsto n-i\ .
\end{equation}
We thus obtain a very simple result: if $\lambda>0$, $g$
flows to the identity component; if $\lambda<0$, the D-brane flows to 
the component where $g$ inverts the order of the eigenvalues of
$t^{\alpha}$. 

In the torus picture (for $k=1$), the identity component corresponds to
the D0-branes. This is what we expect: if the size increases 
($\lambda >0$) beyond the self-dual radius, the D0-branes are the
lightest branes and a generic brane will flow to one of them. If the
size decreases ($\lambda<0$), the physical intuition is less clear,
because there is a B-field on the torus which complicates things. The
torus branes which are described by the inverse ordering of the
eigenvectors correspond to the longest element $w_{0}$ in the Weyl
group.\footnote{Here 'long' refers to the standard length which is the
minimal number of reflections at simple roots needed to write $w_{0}$,
or, in terms of permutations, the minimal number of transpositions of
neighbouring elements.} Its absolute length (minimal number of
reflections, or minimal number of transpositions) is given by 
$\lfloor \frac{n-1}{2} \rfloor$ which 
gives us the dimension of the D-brane on the torus. In the example of
$SU (3)$, the branes which are stable under a perturbation with
$\lambda <0$ are thus D1-branes.
\medskip

So far we have restricted our discussion to a generic perturbation
$t^{\alpha}$. It is clear that there are special directions
$t^{\alpha}$ for which the bulk perturbation breaks less symmetry. If
two or more eigenvalues of $t^{\alpha}$ coincide, one observes
from~\eqref{hstable} that there are more directions $h_{ij}$ which are
unaffected by the flow ($\dot{h}_{ij}=0$), \ie the dimensions of the
moduli spaces of fixed points can grow beyond $n-1$.

For other bulk perturbations $J^{\alpha}\bar{J}^{\bar{\alpha}}$ with
$t^{\bar{\alpha}}\not=t^{\alpha}$, the discussion is very similar to
the one above. Assume that 
$t^{\bar{\alpha}}=\bar{h} (\sum \bar{\tau}_{i} P_{i}) \bar{h}^{-1}$
with eigenvalues $\bar{\tau}_{1}<\dotsb <\bar{\tau}_{n}$. Then the
arguments above apply if we replace $g$ by $\hat{g}=g\bar{h}$. If the
level is $1$, we again have an interpretation in terms of a torus in
$SU(n)$ which is obtained from the Cartan torus by translation by
$\bar{h}^{-1}$ from the right.  
\smallskip

For large values of the level $k$, we can -- as in the $SU (2)$ case
in section~4.2 -- interpret the perturbation as a deformation of the
metric, the B-field and the dilaton on the group
(see~\cite{Forste:2003km}). One would then expect that the group
values to which the branes flow are again characterised by the
property that they maximise the dilaton; it would be interesting to
check this directly.

\section{Conclusions}

In this paper we have studied the interplay between open and closed
string moduli on the disc. In particular, we have shown that an
exactly marginal closed string perturbation (that describes the change
of a closed string modulus) may cease to be exactly marginal in the
presence of a D-brane. If this is the case, the bulk operator induces
a RG flow on the boundary. The end-point of the RG flow is a D-brane
that is conformal in the perturbed bulk theory. We have illustrated
this phenomenon with the example of the free boson theory at $c=1$, 
and with current-current deformations of WZW models.    
\smallskip

It would be interesting to analyse similar phenomena in a
time-dependent string theory context. Suppose, for example, that we
deform the bulk theory of some D-brane string background
infinitesimally so that the D-brane is no longer conformal.
One would then expect that the background evolves in a time
dependent process towards a configuration in which the D-brane is
again conformal.  Neglecting closed string radiation, time dependence is
essentially incorporated by substituting the first order derivatives
in the RG equations by second order time derivatives (see \eg
\cite{Freedman:2005wx,Graham:2006gc}). Since the models we considered
are compact, unlike the situation studied in~\cite{Graham:2006gc}
there is no open string radiation that could escape to
infinity. In particular, there is  
therefore no dissipation and the model will undergo eternal
oscillations. It would be interesting to study the effects of closed
string radiation  in the examples we considered above. In particular,
by suitably controlling the bulk deformation $\lambda$, the process
can be made arbitrarily slow .   
\medskip

Our analysis was originally motivated by trying to understand the
interpretation of the obstruction of \cite{Brunner:2006tc}. 
There $N=2$ supersymmetric B-type D-branes on the orbifold
line $T^4/\Zop_4$ of K3 were studied using matrix factorisation and
conformal field theory techniques. It was found that a certain B-type
brane (namely the brane that stretches diagonally across the two
$T^2$s that make up the $T^4$) is obstructed against
changing the relative radii of the two $T^2$s; this could be seen both
from the matrix factorisation point of view, as well as in conformal
field theory. 

The analysis above suggests that upon changing the relative radii the
brane simply readjusts its angle so that it continues to stretch
diagonally across the two tori. From the point of view of conformal
field theory, there is no obstruction in this. The 
obstruction that was observed in the matrix factorisation analysis
only means that the resulting brane breaks the B-type supersymmetry,
as could also be seen in conformal field theory
\cite{Brunner:2006tc}. It would be interesting to understand more
directly in conformal field theory when such a phenomenon may
happen; the relevant condition will probably be related to the
charge constraint of \cite{Brunner:2006tc}.

At least in this example the obstruction therefore does not
`lift' the corresponding bulk modulus. While we have only analysed
the disc amplitude, we do not expect any higher order corrections
since the brane remains supersymmetric (albeit not B-type
supersymmetric). In general, however, one would expect 
that the backreaction of the brane on the background geometry could
lift bulk moduli. This backreaction is however not visible at the disc
level, and one will have to analyse at least the annulus amplitudes in
order to study it in conformal field  theory. It would be very
interesting to find a simple example where this can be analysed
explicitly.

\vspace{1cm}

\centerline{\large \bf Acknowledgements}
\vskip .2cm
This research has been partially supported by the Swiss National
Science Foundation and the Marie Curie network `Constituents,
Fundamental Forces and Symmetries of the Universe'
(MRTN-CT-2004-005104). MRG thanks the Max Planck Institute in Golm for
hospitality while this paper was completed. 
We thank Costas Bachas, Ilka Brunner, Ben Craps,
Stefan F{\"o}rste, Rajesh Gopakumar, Andreas Recknagel, Ingo
Runkel and Johannes Walcher for useful discussions.

\end{document}